\documentclass[a4paper,10pt]{article}
\usepackage[titletoc,toc,title]{appendix}
\usepackage[utf8]{inputenc}

\usepackage[margin=1in]{geometry} 
\usepackage{amsmath,amsthm,amssymb}

\usepackage{url}
\usepackage{verbatim}
\usepackage{graphicx}
\usepackage{dcolumn}
\usepackage{bm}
\usepackage{color} 
\RequirePackage[normalem]{ulem}  
\usepackage{tabularx}
\usepackage{longtable}
\graphicspath{ {images/} }
\usepackage[table,dvipsnames]{xcolor}
\usepackage{multirow}
\usepackage{setspace}

\newtheorem{claim}{Claim}[section]

\title{Further insights into the damping-induced self-recovery phenomenon}
\author{Tejas Kotwal \thanks{Department of Mathematics} , Roshail Gerard \thanks{Department of
Mechanical Engineering} and Ravi Banavar  \thanks{Systems and Control Engineering, Indian Institute of Technology Bombay, Mumbai, India}}
\date{\today}
\begin{document}

\maketitle

\section{Abstract}

In a series of papers \cite{chang2013_1,chang2013_2,chang2013_3,chang2015}, D. E. Chang, et al., proved and experimentally demonstrated a phenomenon in underactuated mechanical systems, 
that they termed ``damping-induced self-recovery". This paper further investigates a few phenomena observed in these demonstrated
experiments and provides additional theoretical interpretation for these. In particular, we present a model for the infinite-dimensional fluid-stool-wheel system, that approximates its dynamics to that of the better understood finite-dimensional case, and 
comment on the effect of the intervening
fluid  on the large amplitude oscillations observed in the bicycle wheel-stool experiment. The latter is interpreted through an energy analysis from which we conclude that the given model is inadequate to explain the high-amplitude oscillations observed in the experiment in 
\cite{chang2014_1}.

\section{Introduction}
The damping-induced self-recovery phenomenon refers to a fundamental property of underactuated mechanical systems: if an unactuated cyclic variable is subject to a viscous damping-like force and the system starts from rest, then the cyclic variable will always return to its initial state as the actuated variables are brought to rest. A popular illustration exhibiting self-recovery is when a person sits on a rotating stool (with damping), holding a wheel whose axis is parallel to the stool's axis. The wheel can be spun and stopped at will by the person (Refer to \cite{chang2014_1} and Fig. \ref{fig:model}). Initially, the system begins from rest; when the person starts spinning the wheel (say anticlockwise), from momentum conservation consideration, the stool begins moving in the expected direction (i.e., clockwise). The wheel is then brought to a halt; the stool then begins a recovery by going back as many revolutions in the reverse direction (i.e., anticlockwise) as traversed before. This phenomenon defies conventional intuition based on well-known conservation laws. Andy Ruina was the first to report this phenomenon in a talk, where he demonstrated a couple of experiments on video \cite{ruina2010}. Independently, Chang et al., \cite{chang2013_1, chang2013_2} showed that in a mechanical system with an unactuated cyclic variable and an associated viscous damping force, a new momentum-like quantity is conserved. When the other actuated variables are brought to rest, the conservation of this momentum leads to asymptotic recovery of the cyclic variable to its initial position. \textit{Boundedness} is another
associated phenomenon, in which the unactuated cyclic variable converges asymptotically to a non-zero value, when the velocity corresponding to the actuated variable is kept constant; this phenomenon can be viewed as a trade off between the energy input due to the spinning wheel and energy lost due to the damping in the stool. The boundedness phenomenon is kind of a dynamic equilibrium. In the experiment discussed, this manifests as the angle of rotation of the stool reaching an upper limit, when the wheel is spinning at a constant speed.

Chang et al. generalize this theory to an infinite-dimensional system in which the interaction of an intermediate fluid is considered \cite{chang2013_3,chang2015}; they show that the fluid layers also display self-recovery, which is confirmed via experiments as well \cite{chang2014_2}. Such a generalization of the model is considered because of the interaction of the fluid in the bearing with the recovery phenomenon of the stool, in the experiment explained above. In this work, we make the following points:
\begin{itemize}
 \item We show that the dynamics of the stool and the wheel in the infinite-dimensional fluid system can be approximated to that of the finite-dimensional case by finding an effective damping constant that takes into account the effect of the viscous fluid on the system.
 \item In addition to the recovery phenomenon described previously, further complex behaviour is observed in the experiments reported in \cite{chang2014_1,chang2014_2}. In particular, the unactuated variable not only approaches its initial state during recovery, but also overshoots and then oscillates about this initial position, eventually reaching it asymptotically. This oscillation phenomenon has not been looked into in the previous efforts, and is one of the points that we investigate as well.
 \item In Chang et al. \cite{chang2014_1}, in the experiment described, the oscillations are of significant amplitude, and this would prompt one to assume that some sort of mechanical `spring-like' energy is being stored, as the stool appears to start moving after the entire system has come to a halt. The question of energy has been touched upon in this work, and we present energy balance equations for the given mechanical system.
\end{itemize}
 The paper unfolds as follows. Initially we present mathematical models for the stool-wheel experiment -
 the first is a finite-dimensional one, and the second one incorporates the intervening fluid (either
 in the bearing or in a tank) using the Navier-Stokes equation for a Newtonian incompressible fluid. Then follows a
 theoretical section that presents a technique to reduce the infinite-dimensional fluid effect to an
 effective damping constant and hence model the overall system in finite dimensions. In the appendix, the derived results are used in conjunction with numerical experiments to validate the expression for the effective damping constant. We then examine the case of oscillations and 
 overshoots, and present plausible explanations to why these occur, and possible sources
 of future investigation. 
 
 \textbf{An intuitive interpretation of recovery:} A brief explanation of the recovery phenomenon was presented by Andy Ruina \cite{ruina2010} which we summarize: Change in angular momentum is equal to the net external torque, but since it is only due to linear viscous damping, we have $\dot{L} = -c\dot{\varphi}$ (where $L$ is the angular momentum, $c$ is the damping constant and $\varphi$ is the angle). If the system's initial and final state is the rest state, then $\Delta L = 0 \implies \Delta \varphi = 0$. Thus the net change in angle has to be zero, implying that recovery must occur. We build upon this simple explanation to describe why the stool switches direction upon braking the wheel to a halt (when damping is present). The braking of the wheel results in an impulse transmitted to the stool, but since damping tends to slow down the stool, this impulse imparted provides a jump in velocity to the stool that overshoots zero. Hence, there is a change in direction and recovery takes places subsequently.

Another interpretation of the damping-induced recovery phenomenon is to understand it via the theory of bifurcations. If the damping constant is treated as a parameter that can be varied, then it is a trivial exercise to show that as the damping is switched from zero to a positive value,  the system changes from neutrally stable to asymptotically stable. This bifurcation is what leads to the existence of the damping-induced recovery state.
 
\section{Mathematical models}

\begin{figure}[tbh]
   \centering
    \includegraphics[width=0.5\textwidth]{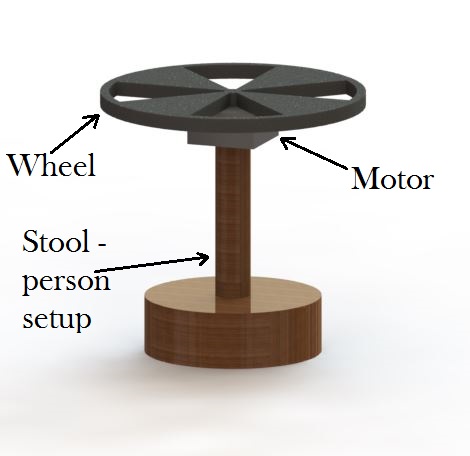}
    \caption{A schematic diagram of the wheel-stool model.}
    \label{fig:model}
\end{figure}

\textbf{Finite-dimensional model:} 
We first analyze a simplified, idealized version of the person with a wheel in hand, sitting on a rotatable stool whose motion is opposed by damping. This is a specific example of the model that Chang, et al. studied \cite{chang2013_1,chang2013_2}. We assume two flat disks, one for the wheel and one for the stool-person mass as shown in Fig. \ref{fig:model}. The stool consists of an internal motor, that actuates the wheel, while the motor-rod-stool setup rotates as one piece (henceforth just called the stool). 
The inertia matrix of the system is given by
\begin{equation}
(m_{ij}) = 
    \begin{pmatrix}
        m_{11} & m_{12} \\
        m_{21} & m_{22}
    \end{pmatrix}
=
    \begin{pmatrix}
        I_{w} & I_{w} \\
        I_{w} & I_{w}+I_{s}
    \end{pmatrix}
,
\end{equation}
where $I_{w}$ and $I_{s}$ are the moments of inertia of the wheel and stool respectively. The kinetic energy of the  system is
\begin{equation}
    K.E.(t) = \frac{1}{2} I_{w} (\dot{\theta}_{w} + \dot{\phi}_{s})^{2} + \frac{1}{2} I_{s} \dot{\phi}_{s}^{2},
\end{equation}
where $\theta_{w}$ denotes the angle rotated by the wheel relative to the stool, and $\phi_{s}$ denotes the angle rotated by the stool relative to the ground frame. Viscous damping is assumed to be present in the rotational motion of the stool, and in \cite{chang2013_1,chang2013_2}, 
this damping force is modelled as $k(\phi_s) \dot{\phi}_s$ with a damping coefficient $k(\phi_s)$ that is a bounded
and piecewise continuous function of the unactuated variable $\phi_s$.
The torque imparted on the wheel by the motor is denoted by $u(t)$.

Since there is no external potential in our system, the Lagrangian only comprises of the total kinetic energy. The Euler-Lagrange equations for the system are given by

\begin{subequations} \label{eq:finite}
    \begin{equation}\label{eq:finiteEulerLagrangeWheel}
        I_{w}\ddot{\phi}_{s} + I_{w}\ddot{\theta}_{w}=u(t)
    \end{equation}
    \begin{equation}\label{eq:finiteEulerLagrangeStool}
        (I_{w}+I_{s})\ddot{\phi}_{s} + I_{w}\ddot{\theta}_{w}=-k(\phi_s)\dot{\phi}_{s}
    \end{equation}
\end{subequations}

Although this model captures the damping-induced boundedness and recovery phenomena, further behaviour that cannot be explained with this model are the overshoot and oscillations, which were observed in experiments \cite{chang2014_1,chang2014_2}. It was suggested that the lubricating fluid inside the bearing of the rotatable stool  could be one of the reasons behind this additional behaviour. As a result, a generalized model that involves the dynamics of the fluid interacting with the stool-wheel setup is analyzed \cite{chang2013_3,chang2015}.

\textbf{Infinite-dimensional fluid model:} The model consists of two concentric infinitely long cylinders with inner radius $R_i$ and outer radius $R_o$. The outer radius is fixed, while the inner cylinder is free to rotate (this is the stool in our case). A motor is installed in the inner cylinder (as described in the previous case, i.e., $\phi_s$), to drive the wheel (i.e., $\theta_w$) attached to it. The annulus region between the two cylinders is filled with an incompressible viscous fluid. Due to symmetry in the $z$ direction, we may regard this as a 2D system in a horizontal plane. The moments of inertia should be considered as per unit depth, because of the infinite length in the $z$ direction. Due to rotational symmetry, the fluid only flows coaxially, i.e, there is no fluid flow in the radial direction. Let $v(r,t)$ denote the tangential velocity of the fluid at radius $r$ and time $t$, where $R_i \leq r \leq R_o$. The subscripts $_t$ and $_r$ denote partial derivatives with respect to time and radius respectively. As described before, $u(t)$ is the driving torque on the wheel imparted by the motor.
The corresponding equations are

\begin{subequations} \label{eq:infinite}
    \begin{equation}
        I_{w}\ddot{\phi}_{s} + I_{w}\ddot{\theta}_{w}=u(t)
    \end{equation}
    \begin{equation} \label{eq:stool_fluid}
        (I_{w}+I_{s})\ddot{\phi}_{s} + I_{w}\ddot{\theta}_{w}=2 \pi \rho \nu R_i \big(R_i v_{r}(R_i,t) - v(R_i,t)\big)
    \end{equation}
    \begin{equation} \label{eq:fluid}
        v_t = \nu (v_{rr} + \frac{v_r}{r} - \frac{v}{r^2})
    \end{equation}
\end{subequations}

where $\nu$ is the kinematic viscosity and $\rho$ is the density of the fluid. The right hand side of Eq. \eqref{eq:stool_fluid} is the torque on the inner cylinder due to stress exerted by the surrounding fluid \cite{landau1959}. Equation \eqref{eq:fluid} is the Navier-Stokes equation for an incompressible viscid fluid in radial coordiantes. The system \eqref{eq:infinite} is an infinite-dimensional fluid system, compared to the previous finite-dimensional system \eqref{eq:finite}. The initial and boundary conditions are given by

\begin{subequations}
    \begin{equation}
        \phi_s(0) = \dot{\phi}_s(0) = \theta_w(0) = \dot{\theta}_w(0) = 0
    \end{equation}
    \begin{equation}
        v(r,0) = 0
    \end{equation}
    \begin{equation} \label{eq:bc1}
        v(R_i,t) = R_i \dot{\phi}_s(t)
    \end{equation}
    \begin{equation} \label{eq:bc2}
        v(R_o,t) = 0
    \end{equation}
\end{subequations}

where Eqs. \eqref{eq:bc1} and \eqref{eq:bc2} are the no-slip boundary conditions for the fluid. The given initial conditions imply that the entire system begins from rest.

\section{Effective damping constant}

In this section, we demonstrate that the dynamics of the infinite-dimensional fluid system can be approximated by that of a finite-dimensional one.

\begin{claim}
	Let the desired trajectory of the wheel be the ramp function (Refer to footnote 1). Then the solution of Eq. \eqref{eq:fluid} satisfies
	\begin{equation*}
	\lim_{t \rightarrow \infty} v(r,t)  =  \lim_{t \rightarrow \infty} R_i \dot{\phi}_s (t) \frac{R_o - r}{R_o - R_i}.
	\end{equation*}
\end{claim}
{\bf Proof}:
We consider an analytical solution to the PDE \eqref{eq:fluid}, obtained by the method of separation of variables. Since one of the boundary conditions is non-homogeneous (and in fact time dependent), we perform a change of variables in order to make it homogeneous \cite{farlow1993partial}. The change of variables is given as

\begin{equation}
    v(r,t) = w(r,t) + R_i \dot{\phi}_s (t) \frac{R_o - r}{R_o - R_i}
\end{equation}

where $w(r,t)$ is transformed the variable. The correction term is taken as a linear interpolation between the two boundary conditions for simplicity. The transformed PDE is given as

\begin{equation}
    w_t - \nu (w_{rr} + \frac{w_r}{r} - \frac{w}{r^2}) = F(r,t)
\end{equation}

where $F(r,t)$ is the driving force of the PDE resulting from the change of variables given by

\begin{equation}
    F(r,t) = - R_i \ddot{\phi}_s (t) \frac{R_o - r}{R_o - R_i} - \nu R_i \dot{\phi}_s (t) \frac{R_o}{r^2(R_o - R_i)}
\end{equation}

The transformed initial and boundary conditions are given by

\begin{subequations}
    \begin{equation}
        w(r,0) = 0
    \end{equation}
    \begin{equation}
        w(R_i,t) = 0
    \end{equation}
    \begin{equation}
        w(R_o,t) = 0
    \end{equation}
\end{subequations}

We assume the solution $w(r,t) = W(r)T(t)$, and obtain the homogeneous solution of the spatial ODE by analyzing the corresponding eigenvalue problem. The eigen solution obtained is given by

\begin{equation}
    W(r) = a J_1 \bigg( \sqrt{\frac{\lambda}{\nu}} r \bigg) + b Y_1 \bigg( \sqrt{\frac{\lambda}{\nu}} r \bigg)
\end{equation}

where $J_1$ and $Y_1$ are the Bessel's functions of the first and second kind respectively, $\lambda$ denotes the eigen value, and $a$ and $b$ are arbitrary constants determined by the boundary conditions. Upon plugging in the boundary conditions, and solving for $\lambda$, we find that the eigen values are positive due to the fact that roots of Bessel's functions are real \cite{abramowitz1964}. Thus we have the following solution for $v(r,t)$

\begin{equation} \label{eq:solution}
    v(r,t) = \underbrace{\sum_{n = 1}^{\infty} \Bigg( J_1 \bigg( \sqrt{\frac{\lambda_n}{\nu}} r \bigg) + k_n Y_1 \bigg( \sqrt{\frac{\lambda_n}{\nu}} r \bigg) \Bigg) T_n (t) }_
    {\text{transient term}}  +  R_i \dot{\phi}_s (t) \frac{R_o - r}{R_o - R_i}
\end{equation}

where $k_n$ and $\lambda_n$ are constants obtained by solving the boundary conditions in the eigen value problem, and $T_n(t)$ is the time component of the solution, which is obtained by solving the corresponding initial value problem. The series given in the right hand side of Eq. \eqref{eq:solution} is considered as a transient term for the PDE \eqref{eq:fluid}, and we show that it is insignificant as $t \rightarrow \infty$. Consider the time component of the solution \eqref{eq:solution}
\[
    T_n(t) = \int_{0}^{t} e^{-\lambda_{n}(t - \tau)} \frac{\int_{R_i}^{R_o} F(r,\tau) W_{n}(r) dr}{\int_{R_i}^{R_o} W_{n}(r)^2 dr} d\tau = e^{-\lambda_{n}t} \int_{0}^{t} e^{\lambda_{n} \tau} (m \ddot{\phi}_{s} (\tau) + l \dot{\phi}_{s} (\tau)) d\tau
\]
where $W_{n}(r)$ is the $n^{th}$ term of the Fourier series, and $m$ and $l$ are constants obtained after evaluating the integrals in the spatial variable. For our case, plug in $\dot{\theta}_{w}(t)$ equal to the desired ramp trajectory (Refer to Section 5 for an explicit description) in the damping-induced momentum equation \cite{chang2013_1}. Upon solving for $\dot{\phi}_{s}(t)$, we see that it is exponentially decaying, and therefore $T_n(t) \rightarrow 0$ as $t \rightarrow \infty$ for all $n$.\footnote{In fact, the result holds for any wheel trajectory that grows polynomially for finite time before eventually coming to rest.} Thus, we have the result.

$\Box.$

In the limit where the transient term is insignificant, the solution \eqref{eq:solution} is plugged in Eq. \eqref{eq:stool_fluid} and simplified as follows

\begin{equation}
    (I_{w}+I_{s})\ddot{\phi}_{s} + I_{w}\ddot{\theta}_{w} = - \frac{2 \pi \rho \nu R_o R_i^2}{R_o - R_i} \dot{\phi}_s
\end{equation}

We would like to emphasize here that we have found an {\it effective damping constant} $k_{eff}$ for the infinite-dimensional fluid system \eqref{eq:infinite} that approximates its dynamics to that of a finite dimension case \eqref{eq:finite}. The effective damping constant is given by

\begin{equation} \label{eq:effective}
    k_{eff} = \frac{2 \pi \rho \nu R_o R_i^2}{R_o - R_i}
\end{equation}
We revisit this effective damping constant and its validity in the appendix.

\section{Oscillations and energy description}

In order to render the wheel dynamics linear, a technique called feedback linearization is adopted, 
wherein a new input $\tau$ is defined as follows
\begin{equation}
    u(t) = \bigg( \frac{I_w}{I_w + I_s} \bigg) \big( I_s \tau(t) + 2 \pi \rho \nu R_i \big(R_i v_{r}(R_i,t) - v(R_i,t)\big) \big)
\end{equation}
 The linearized wheel equations are now $\ddot{\theta}_w = \tau(t)$. If $\theta^d_w(t)$ is the desired trajectory of the wheel, a PD control law for following this desired trajectory is designed as follows 
\begin{equation} \label{eq:control_eq}
    \tau(t) = \ddot{\theta}^d_w(t) + c_1 \big(\dot{\theta}^d_w(t) - \dot{\theta}_w\big) + c_0 \big(\theta^d_w(t) - \theta_w\big)
\end{equation}
where $c_0$ and $c_1$ are the proportional and differential gains respectively, chosen to achieve a desired response of the 
wheel trajectory.

\begin{figure}[tbh]
   \centering
    \includegraphics[width=0.6\textwidth]{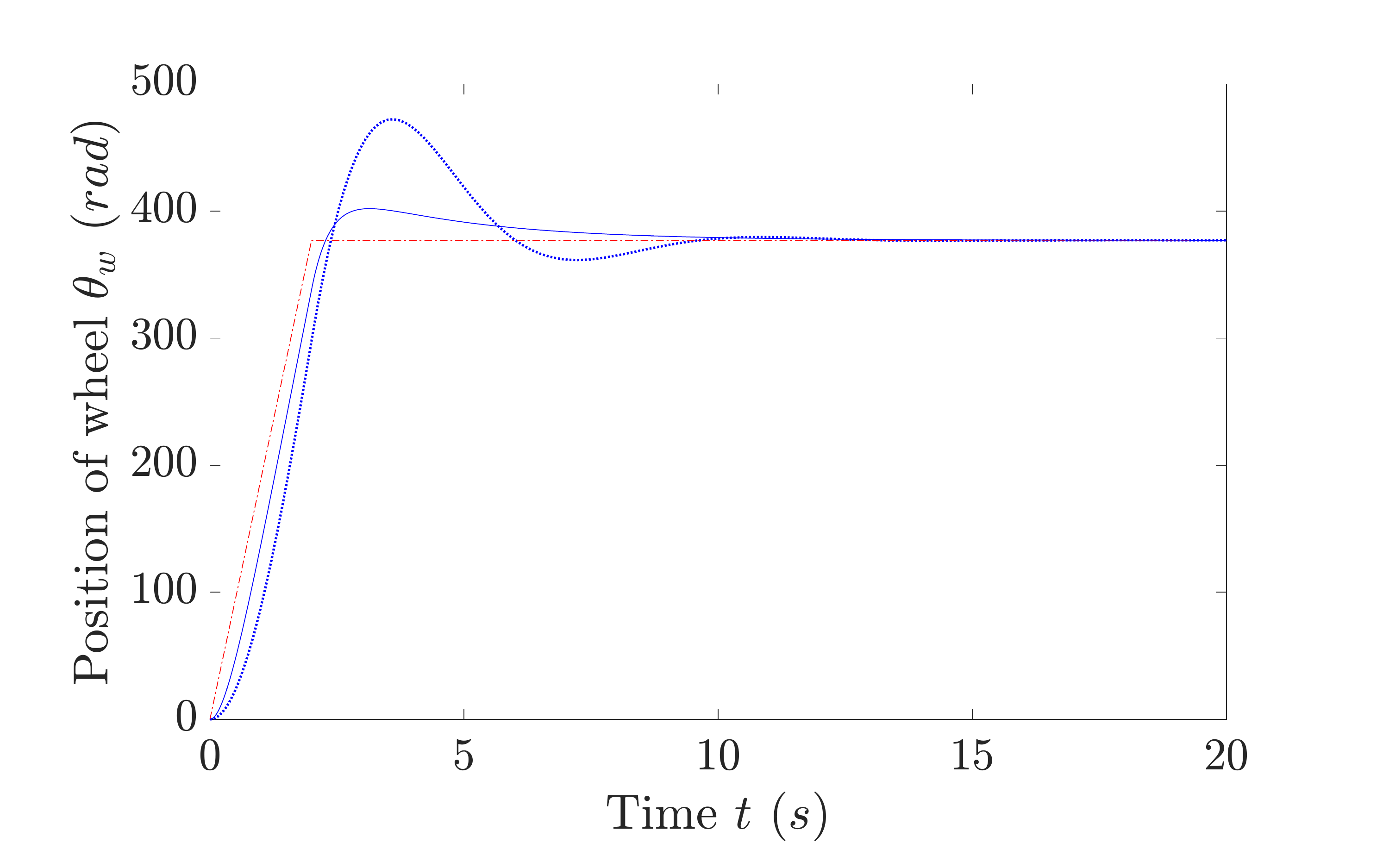}
    \caption{Trajectory of the wheel with control parameters $c_0 = 1$ and $c_1 = 3$ (solid blue line; this is a case of overshoot) and $c_0 = 1$ and $c_1 = 1$ (dotted blue line; this is a case of oscillations). The red dashed line denotes the desired trajectory, and the overshoot of the wheel may be seen at $t = 2 s$.}
    \label{fig:overshoot_wheel}
\end{figure}

To model the recovery and/or boundedness demonstrated in the experiments, we require the wheel to be driven from rest to constant velocity, and then abruptly braked to a stop. Such a desired trajectory of the wheel is given by a ramp function, $\theta^d_w(t) = (1/2) \big( \dot{\theta}_{steady} \big( |t| - |t - t_{stop} | \big) + t_{stop} \big) $ where $\dot{\theta}_{steady}$ is the desired constant velocity of the wheel, and $t_{stop}$ is the time at which the wheel is instantaneously brought to rest (Refer to Fig. \ref{fig:overshoot_wheel}). By appropriately tuning the values of $c_{0}$ and $c_{1}$, we can mimic the desired trajectory. This tuning is important in accounting for the presence of oscillations, or lack thereof, as we shall see shortly.

\begin{figure}[tbh]
   \centering
    \includegraphics[width=0.6\textwidth]{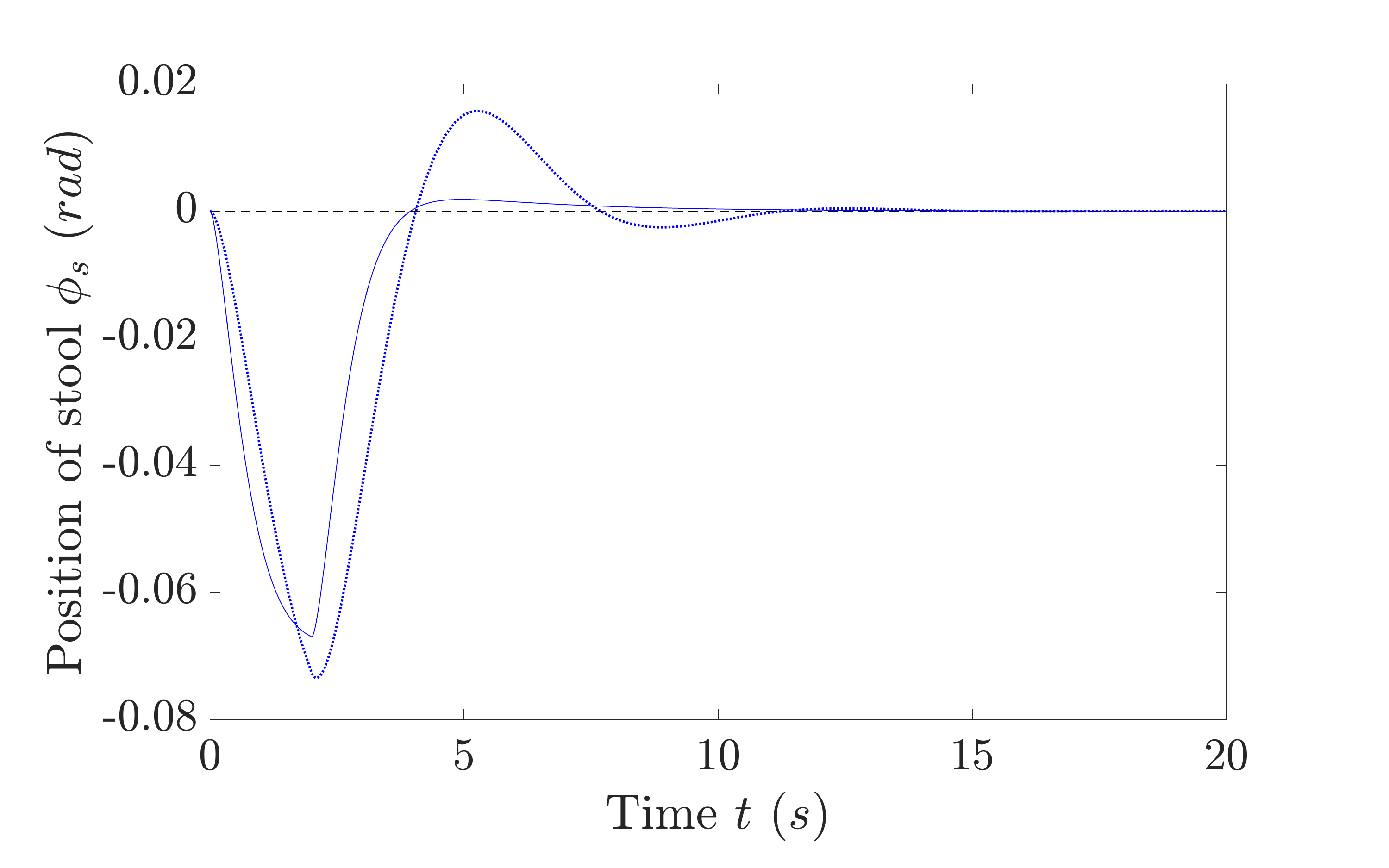}
    \caption{Trajectory of the stool under the influence of the wheel trajectory shown in Fig. \ref{fig:overshoot_wheel}. The black dashed line denotes the zero position of the stool, and the overshoot/oscillations of the stool may be seen after it recovers (Note the slight delay when compared to the overshoot of the wheel; this is due to the time constant of the system).}
    \label{fig:overshoot_stool}
\end{figure}

Oscillations of the unactuated variable have been reported in a few experiments on the damping induced self-recovery phenomenon \cite{chang2015,chang2014_1,ruina2010,chang2014_2}. In our opinion, the cause of these oscillations has not been adequately identified in previous works.
We believe that a seemingly trivial, but important source of these oscillations, is the nature of the control law used on the wheel.
 It is observed that if oscillations are induced in the wheel via the control law, then the stool also mimics these oscillations, with a slight time lag due to the damping. This type of oscillation can be produced in both, the finite-dimensional as well as the infinite-dimensional system. Refer to Figs. \ref{fig:overshoot_wheel}, \ref{fig:overshoot_stool} for examples via simulations. We demonstrate two cases -- first, in which there is only overshoot and second, in which there are oscillations. We would like to emphasize that oscillations of this kind are a direct consequence of the actuation of the wheel. It is likely that the oscillations in \cite{chang2015,chang2014_2} are of this kind.

However, the oscillations in Ref. \cite{chang2014_1} occur even though the wheel is stationary with respect to the stool; in fact, at the extreme positions of the oscillation, it appears as though the entire system (stool and wheel) begins moving again from a complete state of rest. That is, it appears that there is some kind of mechanical `spring-like' energy being stored at these extreme positions. These oscillations are clearly not of the type mentioned above, as they are not a direct result of the actuation of the wheel. Additionally, the observed oscillations in the stool are of high amplitude whereas those in the wheel, if any, appear to be of small amplitude. This is contradictory to what is observed in the previous type of oscillations (seen in Figs. 10 and 11) as the inertia of the stool-wheel system is greater than the inerita of the wheel only.
A simple order of magnitude analysis shows that a more complex model is probably required to account for this phenomenon. In order to demonstrate this, we first consider the energy dynamics of the system.

\textbf{Energy description for the finite-dimensional system:} In order to derive the energy equations of the finite dimensional system, \eqref{eq:finiteEulerLagrangeStool} is multiplied by $\dot{\phi}_{s}$, and \eqref{eq:finiteEulerLagrangeWheel} with $\dot{\theta}_{w}$, to yield

\begin{subequations} 
    \begin{equation}\label{eq:EulerLagrangeVel_Stool}
    (I_{w}+I_{s})\ddot{\phi}_{s}\dot{\phi}_{s} + I_{w}\ddot{\theta}_{w}\dot{\phi}_{s}=-k(\phi_s)\dot{\phi}_{s}\dot{\phi}_{s}
    \end{equation}
      \begin{equation}\label{eq:EulerLagrangeVel_Wheel}
    I_{w}\ddot{\phi}_{s}\dot{\theta}_{w} + I_{w}\ddot{\theta}_{w}\dot{\theta}_{w}=u\dot{\theta}_{w}
    \end{equation}
\end{subequations}

Equations \eqref{eq:EulerLagrangeVel_Stool} and \eqref{eq:EulerLagrangeVel_Wheel} are integrated, and then summed up, resulting in the following energy balance (in the ground frame) 

\begin{equation} \label{eq:EnergyBalance}
    \int_{0}^{t} u \dot{\theta}_{w} \ dt= K.E.(t)+\int_{0}^{t} k(\phi_s) \dot{\phi}_{s}^{2} \ dt
\end{equation}

where K.E.(t) is the total kinetic energy of the system. The second term on the right hand side of represents the cumulative energy lost due to damping losses up to time $t$. The energy lost in damping is precisely equal to the work done by the damping force, that is 

\begin{equation} \label{eq:LostEnergy}
    L.E.(t) = \int_{0}^{\phi_s(t)} k(\phi_s) \dot{\phi}_{s} \ d\phi_{s} = \int_{0}^{t} k(\phi_s) \dot{\phi}_{s}^{2} \ dt.
\end{equation}

The term on the left hand side of Eq. \eqref{eq:EnergyBalance} represents the total energy pumped into the system by the motor, up to time $t$. The power input into the system by the motor is the sum of power imparted to the wheel by the actuation force $u(t)$, and the power imparted to the stool by the reaction force. The cumulative input energy (denoted by $I.E.(t)$) is given by

\begin{equation} \label{eq:InputEnergy}
    I.E.(t)= \int_{0}^{t} u (\dot{\theta}_{w}+\dot{\phi}_{s})\ dt+\int_{0}^{t} (-u) \dot{\phi}_{s}\ dt=\int_{0}^{t} u \dot{\theta}_{w} \ dt
\end{equation}

We can simplify the energy balance \eqref{eq:EnergyBalance} as 

\begin{equation} \label{eq:FinalEnergyBalance}
    I.E.(t)=K.E.(t)+L.E.(t)
\end{equation}

\noindent
\textbf{Energy description for the infinite-dimensional fluid system:} The above energy equation can be extended for the infinite dimensional case as follows.

\begin{equation}\label{eq:FluidEnergyBalance}
I.E.(t) = K.E.(t) + \frac{1}{2} \rho \int v^2 dV + \frac{1}{2} \rho \nu \int_{0}^{t} \int \bigg( \frac{\partial v_i}{\partial x_k} + \frac{\partial v_k}{\partial x_i} \bigg)^2 dV dt
\end{equation}

where the second term on the right hand side is the kinetic energy of the fluid, while the third term is the energy lost due to viscous damping \cite{landau1959}.  The velocity of the fluid in cartesian coordinates is given by $\textbf{v} = -v \ sin \theta \ \hat{i} + v \ cos \theta \ \hat{j}$, where $v$ is the tangential velocity of the fluid as discussed previously \eqref{eq:infinite}. We consider the system at a state when the wheel is no longer being actuated and is in a state of rest with respect to the stool. The system can be found in such a state at the extreme position of an oscillation that occurs after the wheel has been brought to rest. This implies that $I.E(t)$ in \eqref{eq:FluidEnergyBalance} is zero and hence, the kinetic energy of the stool can only be exchanged with that of the fluid, or be lost due to damping. We analyze this state with the parameters from table \ref{table:parametervaluesvideo} for the fluid-stool-wheel system. At the extremum of the oscillation, the stool-wheel system is momentarily at rest, and as is clear from \eqref{eq:FluidEnergyBalance}, must be imparted energy from the fluid to start rotating again. The maximum energy that the stool-wheel system could possibly regain is therefore equal to the kinetic energy of the fluid at this extreme position. This leads to an expression \eqref{eq:EnergyBalanceOsc} for the minimum fluid velocity (averaged over the thickness of the annulus) in terms of system constants and the velocity of the stool. 

\begin{equation}\label{eq:EnergyBalanceOsc}
\frac{1}{2} (I_{w}+I_{s}) \dot{\phi}_{s}^2= \frac{1}{2} \Big[\rho \pi \big[(R_{i}+t)^{2}-(R_{i})^{2}\big]h\Big]v_{avg}^{2}
\end{equation}

where $t$ and $h$ are the thickness and height of the bearing respectively, while $v_{avg}$ is the velocity of the fluid averaged over the thickness of the annulus. Even for a stool velocity as low as 1 rpm, the required average fluid velocity is around 11 rpm, which is quite high and unrealistic. Hence, it is likely that a different model is required to address such behaviour.

\begin{figure}[tbh]
   \centering
    \includegraphics[width=0.6\textwidth]{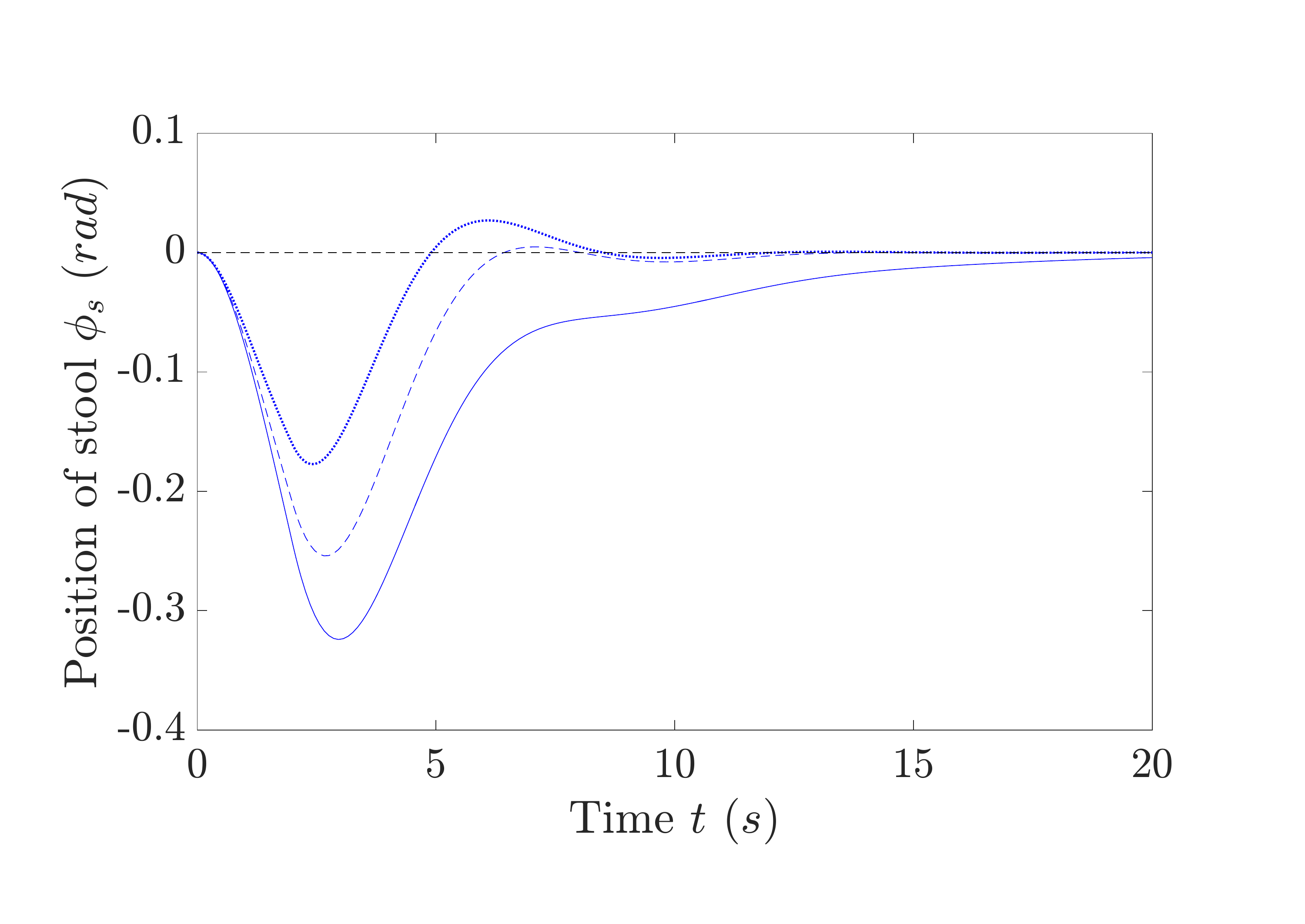}
    \caption{Trajectory of the stool under the influence of the wheel trajectory shown in Fig. \ref{fig:overshoot_wheel} with control parameters $c_0 = 1$ and $c_1 = 1$ and different damping functions $k(\phi_s)$. The black dashed line denotes the zero position of the stool, while the solid blue line denotes the position of the stool when $k(\phi_s) = k/2\pi$, dashed blue line when $k(\phi_s) = k(1 + cos(\phi_s))/2\pi$ and dotted blue line when $k(\phi_s) = 2k(cos(\phi_s))^2/\pi$. The constants chosen are $k = 1$, $I_w = 0.0625$ and $I_s = 0.625$ in model (\ref{eq:finite}).}
    \label{fig:varying_damping}
\end{figure}

We make a short note on how the model could be improved to capture the large-amplitude oscillations in the bicycle wheel-stool experiment shown in \cite{chang2014_1}. Consider the finite dimensional model (\ref{eq:finite}) and empirically choose the damping factor $k(\phi_s)$ (which is a function of $\phi_s$) of the bearing in such a way that the stool closely imitates the high-amplitude oscillations observed in the experiment. This could be one solution to the problem (Refer to Fig. \ref{fig:varying_damping} for an illustration). However, it is unlikely that such a function $k(\phi_s)$ exists because the large oscillations seem to be independent of $\phi_s$ (The initial condition of the stool does not matter as far as the oscillations are concerned). Thus, we suggest that a completely different model is needed to capture this kind of behaviour, and further experimental investigation is necessary.

\begin{table}[tbh]
    \centering
    \begin{tabular}{|c|c|c|}
    \hline
    Name & Value & Units\\\hline
    \hline
    Moment of Inertia of Wheel-Stool System &$1$ & $kg/m^{2}$\\
    Inner diameter of bearing &$0.1$ &$m$\\
    Thickness of bearing &$0.01$ & $m$\\
    Height of bearing &$0.01$ & $m$\\
    Density of Fluid & $1000$ & $kg/m^{3}$\\\hline
    \end{tabular}
    \caption{Parameter values of fluid-wheel-stool system that are representative of experiment in \cite{chang2014_1}}
    \label{table:parametervaluesvideo}
\end{table}

\section{Conclusion}

In this paper, we present certain aspects of the damping-induced self recovery phenomenon that have not been investigated so far in the existing literature. We present a technique to reduce the infinite-dimensional fluid model to the better understood finite dimensional case, by deriving a formula for an effective damping constant. Finally, we present an energy description of the system to give an intuitive understanding of the energy dynamics, and to point out that further experimental and theoretical investigation is necessary to explain the peculiar oscillations found in experiments \cite{chang2014_1,ruina2010}.

\bibliographystyle{unsrt}
\bibliography{sample}

\begin{appendices}
\section{Validation of effective damping constant} 
With the computed, effective damping constant $k_{eff}$ \eqref{eq:effective} of the infinite-dimensional fluid system \eqref{eq:infinite}, we find the angle at which boundedness is attained, given by

\begin{equation} \label{eq:boundedness}
    \phi_s^{*} = \frac{-I_w v_w (R_o - R_i)}{2 \pi \rho \nu R_o R_i^2}
\end{equation}

We now present some numerical experiments by solving the PDE-system \eqref{eq:infinite} using the method of lines. Equation \eqref{eq:effective} is then verified by comparing the 
numerical values obtained using formula \eqref{eq:boundedness} and the numerical solutions of the original 
PDE. It is observed that the error tends to zero as $R_o / R_i$ approaches 1, independent of the value of $R_i$. The material constants used for the simulation, and for the theoretical calculation are tabulated below

\begin{table}[tbh]
    \centering
    \begin{tabular}{|c|c|c|c|c|c|}
    \hline
    \multirow{2}{4em}{$R_i$ (cm)} & \multirow{2}{4em}{$R_o$ (cm)} & \multirow{2}{6em}{$\frac{(R_o-R_i)}{R_i} \times 100$} & With the PDE & With $k_{eff}$ & \multirow{2}{4em}{\% Error}\\
     &  &  & Angle (rad) & Angle (rad) & \\\hline
    \hline
    13.5	&13.51	&0.07	&6.16	&6.16	&0.00\\
    13.5	&13.68	&1.33	&108.7	&109.5	&0.74\\
    13.5	&13.75	&1.85	&149.8	&151.3	&1.00\\
    13.5	&14	&3.70	&291.7	&297.1	&1.85\\
    13.5	&14.5	&7.41	&553.8	&573.7	&3.59\\
    13.5	&15	&11.11	&790.3	&831.9	&5.26\\
    13.5	&15.5	&14.81	&1004	&1073	&6.87\\
    13.5	&20	&48.15	&2265	&2704	&19.38\\
    13.5	&27	&100	&3123	&4160	&33.21\\\hline
    \multicolumn{6}{|c|}{}\\\hline					
    27	&27.02	&0.07	&1.54	&1.54	&0.00\\
    27	&27.36	&1.33	&27.19	&27.37	&0.66\\
    27	&27.5	&1.85	&37.47	&37.81	&0.91\\
    27	&28	    &3.70	&72.96	&74.28	&1.81\\
    27	&29	&7.41	&138.5	&143.4	&3.54\\
    27	&30	&11.11	&197.6	&208	&5.26\\
    27	&31	&14.81	&251.1	&268.4	&6.89\\
    27	&40	&48.15	&566.3	&675.9	&19.35\\
    27	&54	&100	&780.8	&1039.9	&33.18\\\hline
    \end{tabular}
    \caption{A comparison of the angles at which boundedness occurs, i.e., between the numerical simulations and the theoretically derived formula \eqref{eq:boundedness}, for different values of $(R_o - R_i)/R_i$.}
\end{table}

\begin{table}[tbh]
    \centering
    \begin{tabular}{|c|c|c|}
    \hline
    Name & Value & Units\\\hline
    \hline
    Moment of Inertia of Wheel & $6 \times 10^{-3}$ & $kg/m^{2}$\\
    Moment of Inertia of Stool & $1.96$ & $kg/m^{2}$\\
    Kinematic Viscosity of Fluid & $1.17 \times 10^{-6}$ & $m^{2}/s$\\
    Density of Fluid & $1.0147 \times 10^{3}$ & $kg/m^{3}$\\
    Steady State Velocity of Wheel & $60 \pi$ & $rad/s$\\
    Proportional Control Parameter &1 & -\\
    Derivative Control Parameter &100 & -\\\hline
    \end{tabular}
    \caption{Parameter constants used for the simulations.}
    \label{constantvalues}
\end{table}

\end{appendices}

\end{document}